\documentclass[aps,prl,twocolumn,groupedaddress,showpacs]{revtex4}

\usepackage{graphicx}
\usepackage{amssymb}
\usepackage{times}

\begin{document}

\title{Cracks in rubber under tension exceed the shear wave speed}

\author{Paul J. Petersan}
\email[]{petersan@chaos.utexas.edu}
\author{Robert D. Deegan}
\author{M. Marder}
\author{Harry L. Swinney}
\affiliation{Center for Nonlinear Dynamics and Department of
                Physics,
                The University of Texas at Austin,
                Austin, TX 78712}
\date{\today}

\begin{abstract}
The shear wave speed is an upper limit for the speed of cracks
loaded in tension in linear elastic solids. We have discovered
that in a non-linear material, cracks in tension (Mode I) exceed
this sound speed, and travel in an intersonic range between shear
and longitudinal wave speeds. The experiments are conducted in
highly stretched sheets of rubber; intersonic cracks can be
produced simply by popping a balloon.

\end{abstract}



\maketitle

Cracks advance by consuming the potential energy stored in the
surrounding elastic fields and, thus, one expects that they should
travel no faster than the speed of sound.  While this intuition is
indeed correct, solids support longitudinal, shear, and surface
(Rayleigh) waves, with distinct speeds $c_L>c_s>c_R$,
respectively~\cite{Achenbach.73}, and determining which of these
is the upper limit for crack speeds is a subtle problem.

In the case of linear elastic solids, Stroh was first to argue
that cracks loaded in tension (Mode I) could not exceed the
Rayleigh wave speed $c_R$~\cite{Stroh.57}. The mathematical
analysis of Freund and others~\cite{Freund.90,Broberg.99} showed
that the energy consumed by a crack diverges as its speed
approaches $c_R$, suggesting that cracks cannot travel faster than
this speed. However, geological field measurements of fault
motion~\cite{Heaton.90}, computer
simulations~\cite{Andrews.76,Gao.01}, and laboratory
experiments~\cite{Rosakis.02} have established that cracks loaded
in shear (Mode II) can exceed $c_s$ (and hence $c_R$) and reach
speeds close to $c_L$.  Dynamic fracture theory was extended to
include these ``intersonic''
cracks~\cite{Burridge.79,Broberg.99,Rice.00b}. However, even when
revisited in light of the arguments that allowed intersonic cracks
in shear-loaded samples, linear elastic fracture mechanics forbids
intersonic cracks loaded in tension~\cite{Broberg.99}.

In the case of elastic materials subjected to large deformations,
it is natural to assume that the crack speed is bounded above by
the shear wave speed.  Recent numerical simulations have found
mode I cracks running faster than this wave
speed~\cite{Buehler.03}. Here we present experimental results
which show crack speeds faster than the shear wave speed occur in
popping rubber.

\emph{Experiment.} Our experiments are conducted with sheets of
natural latex rubber 0.15~mm thick purchased from McMaster-Carr,
stretched biaxially in a tensile testing machine, and punctured by
pricking the sheet with a needle. The samples are rectangular,
12.7~cm wide and 27.9~-~34.3~cm long, depending on the expected
maximum extension, and their perimeter is divided into tabs 3.0~cm
wide that serve as gripping points for the testing apparatus.  A
square grid is drawn on the sample to monitor the magnitude and
homogeneity of extension as the sample is stretched.

\begin{figure}[h!tb]
\includegraphics[width=3.25in]{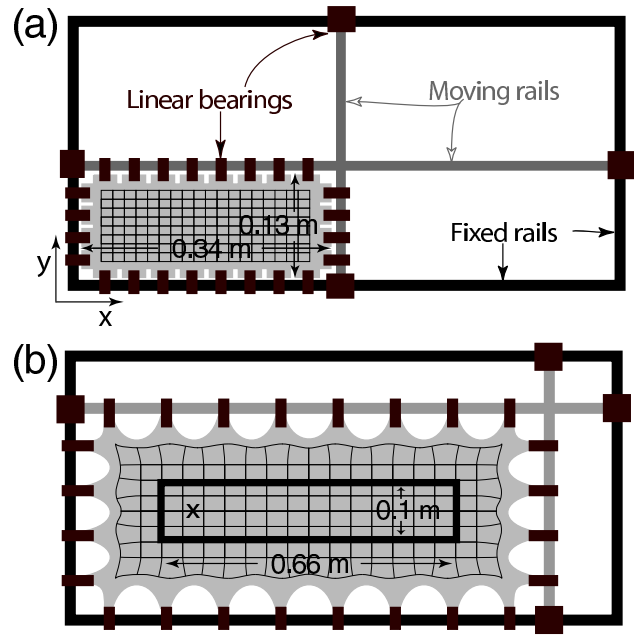}
\caption{\label{fig:apparatus} Apparatus for creating biaxial
extensions in a thin rubber sheets. (a) Initial state of the
sample loaded in the test apparatus.  (b) Final state of the
sample after extension.  The dark rectangle is a rigid frame
attached to the sample and the `$\times$' marks the crack
initiation.}
\end{figure}

The apparatus consists of two fixed and two mobile linear guide
rails (see Fig.~\ref{fig:apparatus}(a)).  The mobile rails move
independently and orthogonally.  Multiple sliding clamps that grip
individual tabs on the sample are mounted on each rail. The sheet
was attached to these clamps and stretched to an extension state
($\lambda_x$,$\lambda_y$) such that $\lambda_y>\lambda_x$, where
$\lambda_i$ is the ratio of the stretched to original length along
the $i$ direction, and $x$ and $y$ are the long and short
directions, respectively. As the rails are moved apart, the sheet
expands and the clamps separate by sliding along the rail, yet
remain equidistant due to the elastic coupling to each other
through the sheet.

After reaching the desired extension state, the sheet is clamped
between two steel rectangular frames (10.2~cm $\times$
66~cm)(Fig.~\ref{fig:apparatus}(b)).  Then the extension state of
the sheet is fixed and determined exclusively by these clamping
frames. The frames fix the total energy available to the crack,
and since the aspect ratio is small, the crack will consume equal
amounts of energy per crack length~\cite{Rice.67}.

\begin{figure}[h!tb]
\includegraphics[width=3.25in]{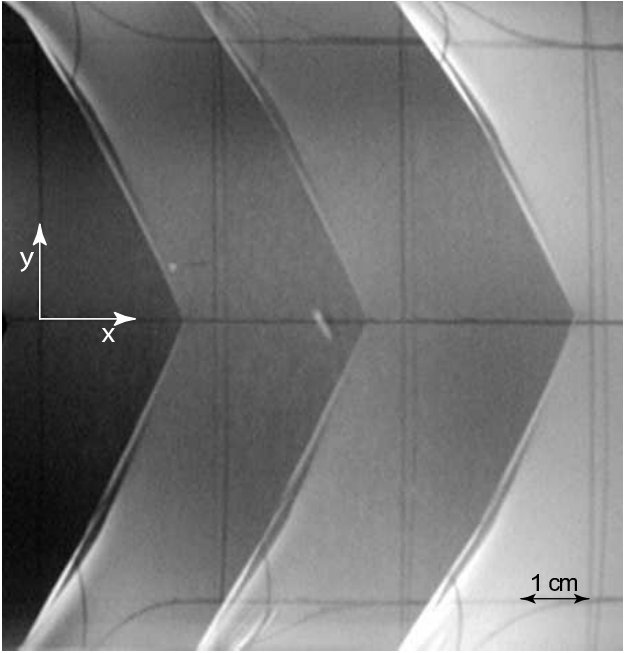}
\caption{\label{fig:crackphoto} Multiple exposure photograph of a
crack in a rubber sample ($\lambda_{x}=2.2, \lambda_{y}=3.2$). The
exposures, separated in time by 500~$\mu$s, are captured on a
single frame with a CCD camera. }
\end{figure}

A crack is initiated by pricking the sheet with a pin at the point
marked '$\times$' in Fig.~\ref{fig:apparatus}(b) and the velocity
is determined photographically.  Along the expected crack path we
position an optical sensor that triggers a strobing flash when the
crack parts the rubber blocking the beam path of the sensor. A CCD
camera captures multiple images from each flash pulse on a single
frame producing an image as shown in Fig.~\ref{fig:crackphoto}. As
reported in~\cite{Deegan.02}, the crack can run straight or wavy
depending on the extension state; irrespective of the crack path,
the velocity of the crack, $v_{c}$, is calculated by the distance
between successive crack tip positions and the time interval
between flashes. Cracks are wavy for  $\lambda_x>2.2$.  These
measurements are reproducible within 5\% from trial to trial.

\begin{figure}[h!tb]
\includegraphics[width=3.25in]{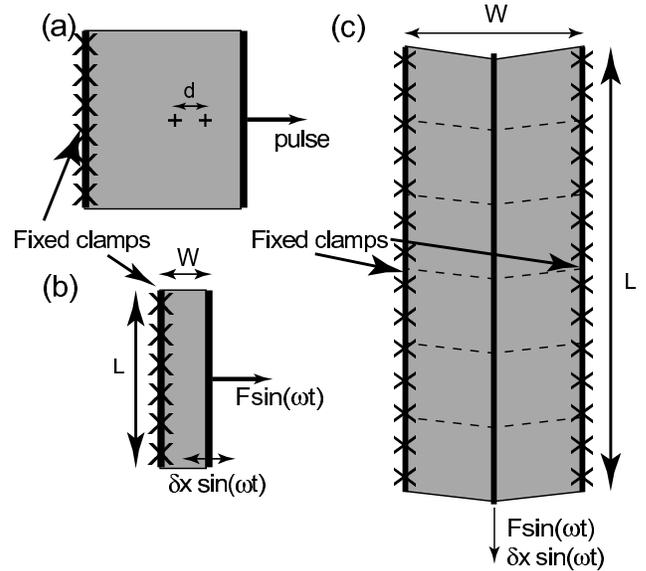}
\caption{\label{fig:speeds} Experimental configurations for
determining sound speeds:  (a) direct measurement of the
longitudinal wave speed using a time-of-flight method; indirect
determination of (b) the longitudinal wave speed and (c) the shear
wave speed from a force-displacement measurement.  In each setup,
the sheet is first brought to the desired state of extension
($\lambda_x,\lambda_y$), the fixed and movable clamps are attached
to the sheet, and the measurements proceed.}
\end{figure}

\emph{Longitudinal wave speed.} We also measure the longitudinal
speed for infinitesimal amplitude waves along the $x$ direction in
rubber as a function of $\lambda_x$ and $\lambda_y$.  In rubber,
sound speeds vary strongly with the degree of extension due to
nonlinear strain-extension relation and large deformations. In
particular, when $\lambda_x\neq\lambda_y$, the sound speed depends
on the direction. We use a time-of-flight method because standard
ultrasonic techniques for measuring sound speeds are difficult to
implement. The sheet is first stretched to an extension state
$\lambda_y>\lambda_x$.  Two record needles (marked by '+' in
Fig.~\ref{fig:speeds}(a)) are placed in contact with the sheet
about 5~cm apart along the $x$ direction.  A bar is attached
across the $y$ direction of the sheet about 5~cm from the first
needle. The bar is attached to a speaker and an in-plane,
$x$--directed perturbation is applied to the sheet by pulsing the
speaker. The signals from the record needles are recorded with a
digitizing oscilloscope, and the longitudinal speed in the $x$
direction is calculated from the time lag between the separate
signals and the known distance between the needles. Note that the
speed is that of a linear wave moving on the stretched material;
i.e., we are measuring the propagation speed of small distortions
around the stretched state.

We have been unable to measure the shear wave speed directly due
to the thinness of the material, which makes it difficult to
excite shear waves.  Instead we measure the appropriate
force-extension curve and calculated the velocity from it.  As a
proof-of-principle, we have calculated in this manner the
longitudinal wave speed.  We measure the force vs. displacement
curve for the configurations depicted in Fig.~\ref{fig:speeds}(b).
The material is stretched to the state ($\lambda_x$,$\lambda_y$),
a long but narrow portion is gripped along the long edges, and one
of the grips is oscillated sinusoidally at $f=25$~Hz in the $x$
direction while the other was fixed. A load cell and accelerometer
attached to the driven clamp measures the applied force $F$ and
acceleration $a$. The displacement $\delta x=a/(2\pi f)^2$ follows
by integration.

Since the measurement is in a strip narrow compared to its length,
departures of the strain in the $y$ direction from the
\emph{stretched} state $\epsilon_{yy}$ is zero almost everywhere.
This state of strain corresponds to the state excited by a
longitudinal wave in a thin sheet: only $\epsilon_{xx}$ and
$\epsilon_{zz}$ are nonzero. Therefore, the force and displacement
measurements in this configuration yield the modulus needed to
calculate $c_L$: $Y=FW/(\delta x L d)$ where $W$ is the distance
between the grips, $L$ is the length of the grips, $d$ is the
thickness of the material in the stretched state, and the
longitudinal wave speed along the $x$ direction is
$c_L=\sqrt{Y/\rho}$, where $\rho=944$~kg/m$^3$ is the mass density
of the sheet.

\begin{figure}[h!tb]
\includegraphics[width=3.25in]{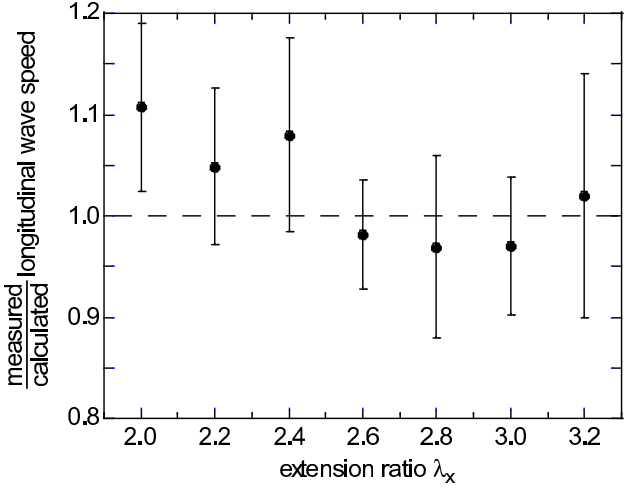}
\caption{\label{fig:ratio} The ratio of measured to calculated
longitudinal wave speed plotted as a function of $\lambda_x$ for
fixed $\lambda_y=3.2$.}
\end{figure}

The ratio of the measured to calculated wave speed is plotted in
Fig.~\ref{fig:ratio} for $\lambda_y=3.2$ and $\lambda_x$ varying
between 2 and 3.2.  For perfect agreement this ratio should be 1.
The data points lie within experimental error of this value,
validating the procedure for calculating the wave speed from the
force-displacement relationship.

\emph{Shear wave speed.}  The configuration in Fig. 3(c) is used
to measure the shear modulus, $G=FW/(4Lt\delta x)$, from which we
calculate the shear wave speed in the $x$ direction, $c_S = \sqrt
{G/\rho}$. A rubber sheet is stretched to $\lambda_y > \lambda_x$,
and a long narrow rubber strip is gripped on opposite edges, as
shown.  A thin bar down is glued to the center line of the strip
and oscillated in the $y$-direction. The result for $G$ obtained
from the force and acceleration measurements then yields $c_s$.

\emph{Material frame.}  It is interesting to examine our data in
the material (Lagrangean) frame rather than the lab (Eulerian)
frame because, as we show below, in the material frame the shear
wave speeds are completely isotropic.  In the material frame,
length is measured in units of a coordinate system that deforms
with the material: for biaxial deformation
$(\lambda_x,\lambda_y)$, the relation between velocity in the
material frame $\mathbf{c}^{(m)}$ and the stretched (or lab) frame
$\mathbf{c}^{(s)}$ is $\mathbf{c}^{(m)}=(c^{(m)}_x,c^{(m)}_y)=(
c^{(s)}_x/\lambda_x, c^{(s)}_y/\lambda_y)$.
%

\begin{figure}[h!tb]
\includegraphics[width=3.25in]{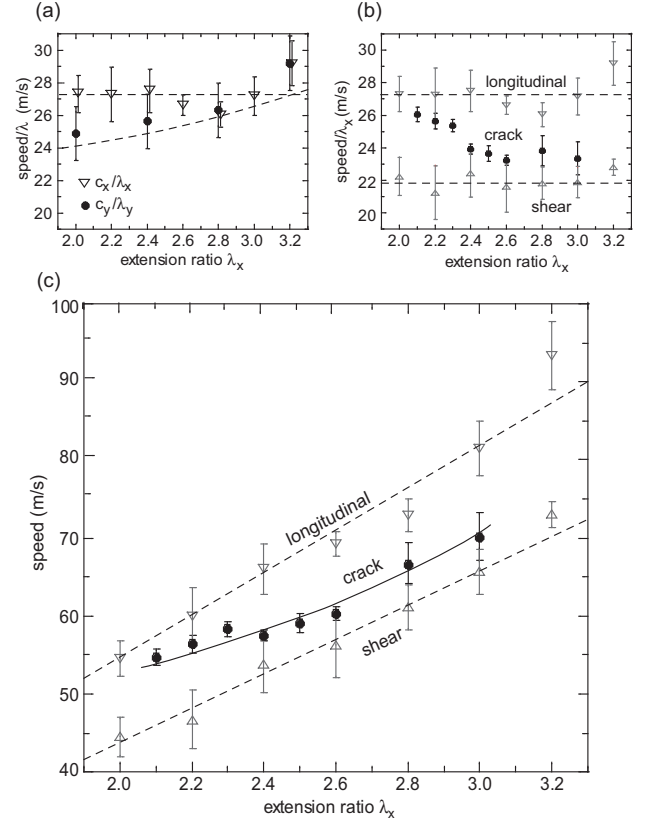}
\caption{\label{fig:crackvel} Comparison of measured longitudinal
and shear wave speeds as a function of $\lambda_x$ at fixed
$\lambda_y=3.2$ with Mooney-Rivlin theory (dashed lines). (a) $x
(\triangledown)$ and $y (\bullet)$ direction longitudinal wave
speeds in the material frame. (b) Longitudinal wave speed
($\triangledown$), crack speed ($\bullet$), and shear wave speed
($\triangle$) in the material frame.  (c) Longitudinal wave speed
($\triangledown$), crack speed ($\bullet$), and shear wave speed
($\triangle$) in the lab frame.}
\end{figure}

\emph{Results.} In Figure~\ref{fig:crackvel} we compare crack
speeds with wave speeds as a function of $\lambda_x$ for
$\lambda_y=3.2$. The data indicate cracks travel 10-20\% faster
than the shear wave speed, but slower than the longitudinal sound
speed. We adopt here the descriptor ``intersonic'' used in studies
of shear loaded cracks, for cracks that travel at speeds between
the longitudinal and shear wave speeds.

\emph{Discussion} There are some natural questions to ask
concerning choices made in Fig.~\ref{fig:crackvel}.  (1) When
rubber is stretched the wave speed is anisotropic.  Why did we
choose just to plot the wave speeds in the $x$ direction?  (2)
Large amplitude deformations typically travel at speeds different
from those of infinitesimal amplitude waves.  Why can we use small
amplitude speeds as a basis for comparison?

The answers to these question stem from the fact that our rubber
samples are described accurately by the Mooney-Rivlin theory,
which we now summarize.  Any continuum theory of deformation that
depends upon relative motion of neighboring mass points can only
involve particle displacements $\mathbf{x}(\mathbf{X})$ through
the strain tensor $E_{\alpha\beta}=(1/2)[\sum_\gamma(\partial
x_\gamma/\partial X_\alpha \partial x_\gamma/\partial
X_\beta)-\delta_{\alpha\beta}]$.  Here $\mathbf{x}(\mathbf{X})$
describes the location in space of a mass point originally at
$\mathbf{X}$. A two--dimensional isotropic theory can only depend
upon the strain tensor through the two invariant quantities
$I_1=E_{xx}+E_{yy}$ and $I_2=E_{xx}E_{yy}-E_{xy}^2$.  Under
biaxial strain $E_{ii}=(\lambda_{i}^2-1)/2$. In the Mooney-Rivlin
Theory~\cite{Mooney.40,Rivlin.48,Shield.66}, the strain energy
density of a thin sheet of rubber is  given by
\begin{equation}
W=a\big{(}I_1+b I_2\big{)} \label{Mooney}
\end{equation}
In the material or Lagrangean frame a calculation of the
longitudinal and shear wave speeds yields:
\begin{eqnarray}
\rho c^{(m)2}_{xl}=a[1+\frac{b}{2}(\lambda_{y}^2-1)]\\
\rho c^{(m)2}_{yl}=a[1+\frac{b}{2}(\lambda_{x}^2-1)]\\
\rho c^{(m)2}_{xs}=a(1-\frac{b}{2})
\end{eqnarray}
To recover the speed in the lab or Eulerian frame
$\mathbf{c}^{(s)}=(c^{(s)}_x,c^{(s)}_y)=( \lambda_x c^{(m)}_x,
\lambda_y c^{(m)}_y)$.  From these equations we see that the
longitudinal and shear wave speeds in the $x$ direction are
independent of $\lambda_x$ and that the $y$ direction longitudinal
wave speed increases monotonically with increasing $\lambda_x$,
both of which we observe. More importantly, $c^{(m)}_s$ is
independent of direction.  With the value of $a = 501$ (m/s)$^2$
and $b = 0.106$, all our sound speed data are well characterized
by these equations, as is shown by the dashed lines in
Fig.~\ref{fig:crackvel}.

Furthermore, in Mooney-Rivlin materials, speeds of infinitesimal
longitudinal and shear waves are exactly equal to the finite
amplitude plane wave speeds. Therefore, although we are measuring
small amplitude wave speeds, these speeds bound the speed of any
finite amplitude plane wave deformation. Thus, it is appropriate
to compare the wave speeds we measured to crack speeds.

Lastly, we note that when a moving object travels in an elastic
medium faster than the wave speed, it creates a shock wave in the
form of a Mach cone.  The wedge-like crack opening typical of
cracks in rubber (shown in Fig.~\ref{fig:crackphoto}) is
strikingly similar to the Mach cone, suggesting that rubber cracks
are shock-like, exceeding some response speed of the medium.
Shock waves are known to form when finite tensile pulses travel in
pre-stretched rubber strips~\cite{Kolsky.69}.

\emph{Conclusion.} In earlier work we found that cracks in rubber
follow oscillating paths when the horizontal extension exceeds a
critical value. We still have no good explanation for this
dynamical transition. However, knowing that the fractures travel
above the shear wave speed will be important in resolving this
question.

\begin{acknowledgments}
\emph{Acknowledgements---}We thank James Rice for encouraging
further investigation of the Mooney-Rivlin theory and Stephan
Bless for lending us the stroboscopic flash used in imaging the
moving crack.  We are grateful for financial support from the
National Science Foundation (DMR-9877044 and DMR-0101030).
\end{acknowledgments}

\end{document}